\documentstyle[aps,epsfig]{revtex}

\begin{document}

\draft

\def\be{\begin{equation}}
\def\ee{\end{equation}}
\def\beqa{\begin{eqnarray}}
\def\eeqa{\end{eqnarray}}
\def\x{\vec{x}}
\def\k{\vec{k}}
\def\e{{\rm e}}
\def\im{{\rm i}}

\title{Spatial correlations in  hexagons generated 
via a Kerr nonlinearity}

\author{Alessandra Gatti$^1$ and  Stefano Mancini$^2$}

\address{$^1$ INFM, Dipartimento di Scienze, 
Universit\'a dell'Insubria, 
Via Valleggio 11, I-22100 Como, Italy \\ 
$^2$ INFM, Dipartimento di Fisica,
Universit\`a di Milano,
Via Celoria 16, I-20133 Milano, Italy} 

\date{\today}

\maketitle

\widetext

\begin{abstract}
We consider the hexagonal pattern forming 
in the cross-section
of an optical beam produced
by a Kerr cavity, and 
we study the quantum correlations characterizing                      
this structure. 
By using arguments related to the symmetry broken by 
the pattern formation, we
identify a complete scenario of six-mode entanglement.
Five independent phase quadratures combinations, connecting
the hexagonal modes, are shown to exhibit sub-shot-noise 
fluctuations. 
By means of a  non-linear quantum calculation technique,
quantum correlations among the mode photon numbers are demonstrated
and calculated. 
\end{abstract}


\section{Introduction}

It is well-known that spatial patterns may arise 
spontaneously in the
transverse cross-section of optical beams as a consequence 
of nonlinear
wave mixing processes \cite{ACA}. Equally fascinating, by maybe 
less known, is the presence
of highly non classical spatial correlations in the beam 
cross-section, 
underlying the classical process of pattern formation.
In optical systems the nonlinearity is associated with the 
simoultaneous 
absorption and emission of a number of photons, a circumstance 
that creates 
entanglement among the waves 
that form the pattern. 
On a macroscopic level, this is at the very origin of non-classical
spatial correlations in the transverse far-field plane \cite{ACA}b.

Among the possible spatial structures that may spontaneously arise
in a beam cross section, the hexagonal pattern is one of the 
most common ones.
Hexagonal structures have been predicted to form,  typically, 
in Kerr or Kerr-like
media in various configurations. These include 
counterpropagating 
waves interacting with a slice of Kerr material  
in a cavityless configuration \cite{FIRTH88}, 
systems with a single feedback mirror \cite{FIRTH90},
and planar resonators filled  with a self-focusing
Kerr medium \cite{FIRTH92}, or with a non-saturable 
absorber \cite{nonsatabs}. 
Experimental observations of this kind of 
optical pattern include counterpropagating beams in sodium 
\cite{GRYN88}, 
and a variety of systems with single feedback 
(see e.g. \cite{singlefeed}).

This paper is devoted to the analysis of the quantum 
properties of the hexagonal pattern,
in the context of a model for a Kerr medium enclosed in 
a planar resonator 
driven by a plane-wave pump beam. This model represents, 
for its simplicity,
a paradigm for optical pattern formation. In fact, 
it was one of the the first models
able to predict a
tranverse modulational instability in an optical 
system \cite{LL87}.
If only one dimension in the transverse plane is 
considered \cite{LL87},
the modulational instability gives origin to a roll 
pattern in the near field immediately out
of the cavity.   
The two dimensional version of the model, 
analysed by  \cite{FIRTH92},
predicts that,
above a suitable threshold of the input intensity, 
a spontaneous breaking of the translational 
symmetry in the transverse plane 
gives rise to
a hexagonal pattern in the near field, which in the 
far field corresponds to six 
bright spots in hexagonal arrangement surrounding a central spot. 

Quantum aspects of the hexagonal pattern in the Kerr cavity  were
first evidenciated  by Grynberg and Lugiato \cite{GL93}. 
Quantum correlations in the 
light intensities of groups of four among the six hexagonal spots  
were predicted by using momentum-energy conservation arguments, 
which did not take into account the effects of dissipation through  
the cavity mirror (hence valid only for a single pass through 
the nonlinear medium).

As in the approach of \cite{GL93} we will 
restrict our analysis to a seven-mode model, valid close to 
the instability threshold.
By using a  quantum calculation technique which takes 
into account the full nonlinearity
of the problem, 
we will be able to demonstrate that the results of \cite{GL93}. 
hold also in the presence of cavity dissipation. Moreover, we will
derive an analytical formula for the fluctuation 
spectrum describing the quantum correlation among mode intensities. 
 
In the second part of the paper we will resort to more traditional
calculation technique, namely to the usual linearization
of the quantum fluctuation dynamics around the classical 
steady-state pattern. By using arguments related to  the  
translational 
symmetry of the model, broken by the pattern formation,
we  shall identify a complete scenario of quantum 
correlation among the phase
quadratures of the six hexagonal modes.

We believe that the six mode entanglement described in this paper, 
besides being likely to be accessible to experiments, 
might be of relevance 
from the point of view 
of quantum information applications.

Section \ref{themodel} introduces the quantum mechanical 
and classical models
describing the self focusing Kerr cavity. Section \ref{nmeno} 
is devoted to
photon number correlations among the hexagonal modes in the 
context of 
a nonlinear model. Section \ref{phquad} investigates in 
general the phase quadrature
correlation among modes. Section \ref{conclusions}
concludes and discusses open questions.

\section{The model} \label{themodel}

In this section we present the quantum mechanical 
counterpart \cite{LUCA92} 
of a well-known semi-classical model \cite{LL87,FIRTH92}, 
which describes 
the dynamics of the
scalar electric field in a cavity filled with an isotropic 
Kerr medium. 

\begin{figure}[t] 
\centerline{\epsfig{figure=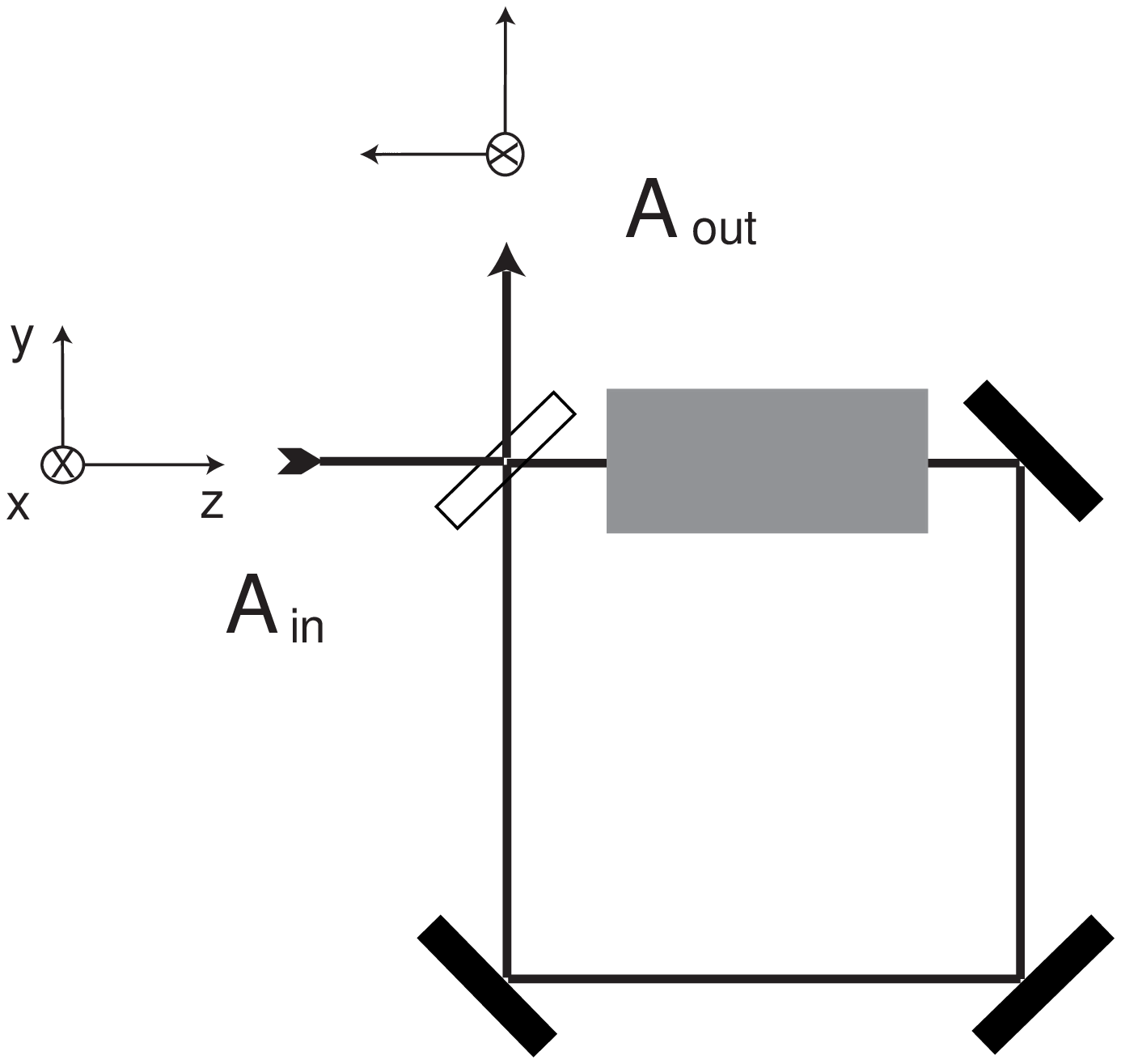,width=10cm}}
\vglue 2.true cm
\caption{
Scheme of a ring cavity containing a Kerr medium (K).
The longitudinal and transverse axis are shown for both 
input and output field.} 
\label{fig1} 
\end{figure}

We consider a one-directional planar cavity (Fig. \ref{fig1}), 
with a single input-output port,   
driven by a coherent, plane-wave,
monochromatic field of frequency $\omega_0$. 
The model is derived in the framework of the  
slowly varying envelope 
and paraxial
approximation, and of the cavity mean field limit \cite{meanfield},
which  allows to neglect the dependence of the field on the
longitudinal coordinate $z$ along the sample. Under these
assumptions only one longitudinal cavity mode is relevant,
precisely the one corresponding to the longitudinal cavity
resonance $\omega_c$ closest to $\omega_0$. We denote by $A(\vec
x, t)$ the intracavity field envelope
operator corresponding to this mode; it depends on 
the transverse space
coordinate $\vec x = (x,y)$ and time $t$, and obey standard
equal-time commutation relations
\begin{equation}
\left[A(\vec x, t), A^\dagger(\vec x', t)\right]
=\delta(\vec x-\vec x')\,.
			\label{equaltimecommutator}
\end{equation}
By adopting a picture where the fast oscillation at the carrier 
frequency $\omega_0$ is
eliminated, the reversible part of the intracavity field dynamics
is governed by  the three 
Hamiltonian terms 
\begin{equation}
H=H_{free}+H_{ext}+H_{int} \; .
			\label{hamiltonian}
\end{equation}
The coupling
due to the Kerr nonlinearity in the medium is accounted for by
\begin{equation}
H_{int}=-\hbar \gamma \frac{g}{2} \int d^2 \vec x 
 \left[A^\dagger(\vec x)\right]^2 
\left[A^{\phantom \dagger}(\vec x)\right]^2
             \label{Hint}
\end{equation}
where the coupling constant $g$ is proportional to the 
third order  $\chi^{(3)}$
susceptibility, and $\gamma$ is the cavity linewidth.
This Hamiltonian describes four-wave-mixing interactions
at a microscopic level, where they correspond to simultaneous 
annihilation and creation of photons in pairs \cite{LUCA92}.
Free propagation in a planar cavity is described, in the paraxial
approximation, by
\begin{equation}
H_{free}=\hbar \gamma \int{d^2 \vec x A^\dagger(\vec x)
\left( \Delta- l_D^2\nabla^2 \right)A(\vec x) }\, ,
\label{freehamiltonian}
\end{equation}
where $\Delta$ is the cavity detuning parameter; the two
dimensional transverse Laplacian $\nabla^2= 
\frac{\partial^2}{\partial x^2} +\frac{\partial^2}{\partial y^2} $ 
models the effect of
diffraction in the paraxial approximation; 
the parameter $l_D \approx
\sqrt{L \lambda/ (2 \pi {\cal T})}$, with $\lambda$ being the light
wavelength, $L$ the total cavity length, and 
${\cal T}$ the transmittivity coefficient of the cavity
input-output mirror, defines the characteristic length 
scale for transverse pattern formation.  
Finally,
the coherent pumping, by a classical
plane-wave driving field of amplitude $E_{in}$, 
is modeled by the term: 
\begin{equation}
H_{ext}=i \hbar \gamma \int{d^2 \vec x \; [E_{in} A^\dagger(\vec x)
-E_{in}^*A(\vec x)]} \, .
\label{externalhamiltonian}
\end{equation}
Dissipation through the cavity mirror can be described 
in the framework of standard cavity input-output formalism 
(see e.g. \cite{GAR}). As a result the
dynamics of the intracavity envelope operator is governed 
by the equation 
\be
\frac{\partial}{\partial t} A(\x,t) = -\gamma  
             \left[\left( 1 + i \Delta -i l_D^2 \nabla^2 \right)
             A(\x,t) 
			-i g A(\x,t)^\dagger A^2(\x,t) -E_{in} \right] 
			+ \sqrt{2\gamma} \delta A_{in} (\x,t) \, ,
				\label{Aeq}
\ee
where the noise operator $\delta A(\x,t)_{in} $ 
represents the vacuum 
fluctuations entering the
cavity input-output mirror. It has zero mean, and obeys 
the free field 
commutation relation
\be
\left[\delta A_{in} (\x,t) \,, \, \delta A_{in}^\dagger(\x',t')  
\right] 
= \delta(\x-\x') \delta(t-t')\,.
				\label{Aincomm}
\ee
Equation~(\ref{Aeq}) must be coupled with the relation  
at the cavity mirror, 
linking the outgoing reflected field
$A_{out} (\x,t)$ with the intracavity and input fields
\be
A_{out} (\x,t)= \sqrt{2\gamma} A (\x,t) -A_{in} (\x,t) \; ,
\ee
where, in our notation, the total input field operator
is given by $A_{in}(\x,t)=\sqrt{\gamma/2} E_{in}  
+ \delta A_{in}(\x,t)$.

The classical counterpart of this model is the well known   
partial differential equation \cite{LL87}
\begin{equation}
				\label{eqLL}
\frac{1}{\gamma}\frac{\partial E}{\partial t}= -E+E_{in} +
i\left(|E|^2-\Delta + l_D^2\nabla^2 E \right)E \,,
\end{equation}
for the envelope of the classical electric field $E$. This equation 
is straightforwardly obtained by its operatorial version 
(\ref{Aeq})  by
dropping the noise term,  and scaling the input 
and intracavity field 
intensities  to the saturation 
photon number $1/g$, i.e. $g|E|^2 \rightarrow 
|E|^2$,  $g|E_{in} |^2 
\rightarrow |E_{in}|^2$.

Since the model has an overall translational symmetry 
in the transverse plane, equation (\ref{eqLL}) admits a transversely 
homogeneous stationary solution $E_{0s}$,
which obeys the well-known cubic steady-state equation \cite{GIBBS76}
\begin{equation} {}
				\label{E0s}
|E_{in} |^2=|E_{0s}|^2\left[1+\left(\Delta
-|E_{0s}|^2\right)^2\right]\,.
\end{equation}
As shown by Lugiato and Lefever \cite{LL87}, 
by increasing the input intensity 
a spontaneous breaking of the translational symmetry occurs, and
the steady state~(\ref{E0s}) becomes unstable with respect to 
perturbations spatially  modulated 
in the transverse plane. As it is common in pattern formation 
processes \cite{ACA} the instability develops
with a definite critical wave number $k_c$,  which characterizes the 
periodicity of the pattern 
immediately above the instability threshold.
The critical point for the onset of the modulational instability is
given by \cite{LL87}
\begin{equation}
\label{critical}
|E_{0s}|^2=1\,,\qquad l_D^2 k_c^2=2-\Delta\,.
\end{equation}
Among the transverse modes on the critical circle $|\k|=k_c$, 
the Kerr nonlinearity
selects a discrete set of six wave vectors \cite{FIRTH92,SCRO94},
so that immediately above the critical point
the near field intensity distribution
consists of a hexagonal lattice of bright spots ,  
while the far field shows a central bright spot 
surrounded by six spots in 
hexagonal arrangement.

\begin{figure}[t]
\centerline{\epsfig{figure=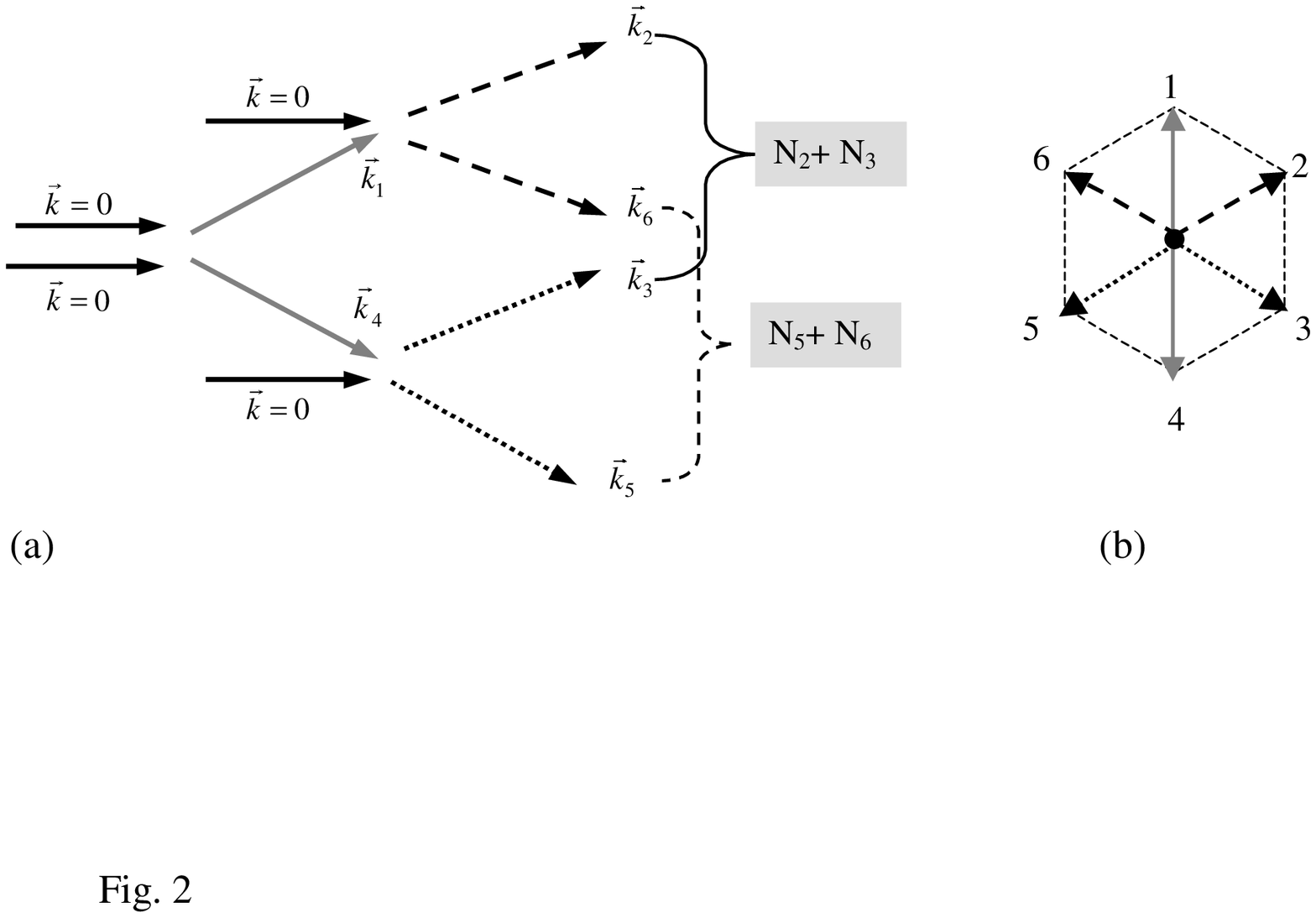,width=14cm}}
\caption{(a)Scheme of the microscopic processes leading to 
hexagonal pattern  generation (see text). (b) Transverse 
wave vectors forming the hexagonal pattern.}
\label{fig2}
\end{figure}

The origin of this particular 
pattern can be understood by considering the microscopic 
processes leading 
to off-axis emission of photons  allowed by energy-momentum 
conservation
\cite{GRYN88th}.
By referring to the  scheme of  figure~\ref{fig2}a,
a first process that preserves the radiation transverse momentum
involves destruction of two photons of the plane-wave input beam,
and creation of two photons propagating off-axis in 
symmetrical directions.
This is the only possible process  in a 1-D version of 
this model \cite{LL87},
and would lead to the formation of a stripe pattern 
in the near field, as 
e.g. predicted by a vectorial model for a self-defocusing Kerr 
medium \cite{GEDDES94,HOYUELOS98},
or in the case of a degenerate optical parametric oscillator 
\cite{ROLLS}.
In this system, however,
the stripe pattern is unstable, because secondary four-wave mixing 
processes can take place. 
By referring to figure~\ref{fig2}, we can e.g. consider destruction 
of one on-axis photon
of the input beam and one photon propagating in direction 1, 
and creation
of two off-axis photons propagating in directions 2 and 6. 
By considering also the symmetric process,
that involves destruction of a photon in mode 4 and one in 
the on-axis mode, 
and creation of two 
photons in mode 3 and 6, respectively, we see that the six 
off-axis hexagonal 
modes are activated. 
Close to the threshold for pattern formation, all the other 
processes allowed
by energy-momentum conservation, that do not involve 
on-axis photons,
can be neglected, because off-axis mode  intensities are much 
weaker than that of the on-axis homogeneous mode, where the pump 
beam is injected.

\section{Photon number correlation in the hexagonal pattern} 
\label{nmeno}

The microscopic processes  which generate the hexagonal pattern
are at the origin of  non-classical correlation among  different 
portions of the far-field cross section. Referring to the example of 
figure~\ref{fig2},
it can be noticed that whenever two photons are detected 
at positions
2 and 3 in the far field, there must be two other photons 
at position 5 and 6,
which suggests the existence  of a  high level  of correlation 
between the sum of photon numbers
$N_2 + N_3$ and $N_5 + N_6$ (the same reasoning can be applied to 
any symmetric group of four spots).

A more quantitative argument for the existence of such a kind of 
intensity correlation in
the hexagonal pattern was given by Grynberg and Lugiato\cite{GL93}. 
They considered a discrete model for the hexagonal pattern 
formation in
Kerr media, which takes into account only the six hexagonal 
modes plus 
the homogeneous mode $\k=0$,
valid close to the threshold for pattern formation, 
where these modes gives 
the most relevant contribution to the system dynamics. 
By considering
a quantization box in the transverse plane of side $b$  
($b \to \infty$ 
at the end of the calculations)
they set
\be
A(\x,t) = \frac{1}{b} a_0 (t) + \frac{1}{b} \sum_{i=1}^{6} a_i (t) 
e^{i \k_i \cdot \x}  \; ,
					\label{expansion}
\ee 
with (see figure \ref{fig2}b) $\k_1= (0,k_c)$, $\k_2=(k_c/2, 
\sqrt{3}/2 k_c)$,
$\k_3= (k_c/2, -\sqrt{3}/2 k_c)$,  
$k_4=-\k_1$, $k_5=-\k_2$,  $k_6=-\k_3$. By inserting this expansion 
into the system Hamiltonian (\ref{hamiltonian})
they found that the combination of mode photon numbers 
\be
N_-= N_i + N_{i \oplus 1}- \left( N_{i \oplus 3}+N_{i \oplus 4} 
\right)\,,  
\qquad 
i=1,\ldots,6\,, 
\qquad
i\oplus j \equiv i+j \;\rm{mod}\; 6
\label{N_}
\ee
with $N_i = a^\dagger_i a_i$,
commutes with the discretized Hamiltonian. 
Thus, in a single pass through 
the nonlinear medium 
the observable $N_-$ is preserved. If there are no photons in the 
hexagonal modes at the crystal entrance, 
perfect correlation is found at the exit between e.g. 
$ N_i + N_{i \oplus 1}$ and
$ N_{i \oplus 3}+N_{i \oplus4}$. However, when the crystal 
is inserted in an open resonator, 
the field is recycled by the cavity and dissipation
through the cavity input/output mirror should be taken into account. 
As a matter of fact, the release
of a photon out of the cavity mirror 
is a random process uncorrelated with
the nonlinear processes taking place in the crystal, 
so that one would expect that cavity dissipation is detrimental for  
this kind of correlation.

However, we will now show that this is not the case.
In order to elucidate this point, we shall 
follow two different techniques. 
The standard one,
which will be employed in the next section, 
makes use of the small quantum noise approximation, and consists of 
linearizing the dynamics of quantum fluctuations 
around a classical steady state. The other one 
consists of trying to solve the non-linear
Langevin equations that govern 
the dynamics of the photon number operators of the hexagonal modes. 
These equations take the general form (see e.g. \cite{GAR})
\be
\frac {d} {d t} N_i = -2 \gamma N_i + \frac {1}{i \hbar }
\left[N_i ,\, H \right]+ \sqrt{2\gamma}
\left[ a_i^\dagger a_i^{in} + a_i^{\dagger \, in } a_i \right]
				\label{dndt}
\ee 
with $a_i^{in}$ being noise operators modelling the vacuum input 
fluctuations on $i$-th mode. They obey standard
free-field commutation rules 
\be
\left[ a_i^{in}  (t), a_j^{\dagger \, in } (t') \right]
=\delta_{i,j} 
\delta(t-t') 
					\label{incomm}
\ee
and have the noticeable commutator with a generic 
intracavity operator 
${\cal O}$ \cite{GAR}
\be
\left[ {\cal O}(t), a_i^{in}  (t') \right] = \sqrt{2\gamma}
\theta(t-t') \left[ {\cal O}(t), a_i (t') \right] \; ,
							\label{causality}
\ee
where $\theta$ is the step function.
Equation (\ref{causality}) expresses the fact that 
intracavity operators at time $t$  
commute with the input operator at a later
time $t'$ due to causality. 
The discretized version of Hamiltonian (\ref{Hint}),
obtained by means of the expansion (\ref{expansion}), 
results \cite{GL93}
\begin{equation}  
H_{\rm int}=H_{\rm SPM}+H_{\rm CPM}+H_{\rm FWM}^{(1)}
+H_{\rm FWM}^{(2)}+H_{\rm FWM}^{(3)}\,,
				\label{DHINT}
\end{equation}
where $H_{\rm SPM}$ describes self-phase modulation
\begin{equation}
\label{DHSPM}
H_{\rm SPM}=-\hbar \gamma \frac{g}{2}\sum_{j=0}^{6}\left(a_j^{\dag}
\right)^2\left(a_j\right)^2\,,
\end{equation}
while $H_{\rm CPM}$ describes cross-phase modulation
\begin{equation}
\label{DHCPM}
H_{\rm CPM}=-2 \hbar\gamma g \sum_{i<j} 
a_i^{\dag}a_i a_j^{\dag}a_j\,,
\end{equation}
and $H_{\rm FWM}$ describes four-wave mixing
\begin{equation}
\label{DHFWM1}
H_{\rm FWM}^{(1)}=-\hbar\gamma g \left[\sum_{j=1}^{3} 
a_0^2 a^{\dag}_ja^{\dag}_{j\oplus 3}+{\rm
h.c.}\right]\,,
\end{equation}
\begin{equation}
\label{DHFWM2}
H_{\rm FWM}^{(2)}=-2\hbar \gamma g \left[\sum_{i<j}^{3}
a_i a_{i\oplus 3} a^{\dag}_j a^{\dag}_{j\oplus 3}+{\rm h.c.}
\right]\,,
\end{equation}
\begin{equation}
\label{DHFWM3}
H_{\rm FWM}^{(3)}=-2\hbar \gamma g  \left[\sum_{j=1}^{6}
a_0 a_{j} a^{\dag}_{j\oplus 1} a^{\dag}_{j\oplus 5}+{\rm h.c.}
\right]\,.
\end{equation}
Other terms of the Hamiltonian commute with the photon number 
operators.

The second term at RHS of the dynamical equations (\ref{dndt}) 
is therefore
a complicated expression, involving 
third order products of mode operators, so that finding an 
explicit solution seems
unlikely. However, when considering the combination of mode 
photon numbers 
$N_-$ given by 
Eq.(\ref{N_}) the dynamics drastically simplify:
\beqa
\frac {d} {d t} N_- &=& -2 \gamma N_- - \frac {i}{\hbar }
\left[N_- ,\, H \right]+ \sqrt{2\gamma}
G_- \,,
\\ 
&=& -2 \gamma N_- + \sqrt{2\gamma} G_-\,, 
								 \label{dN-dt} 
\eeqa
where we used the fact that $N_-$ commutes with the system 
Hamiltonian, 
and the operator $G_-$ is defined as:
\beqa
G_-&=& a_i^\dagger a_i^{in} + a_i^{\dagger \, in } a_i 
 + a_{i\oplus 1 }^\dagger a_{i\oplus 1 }^{in} 
 + a_{i\oplus 1 }^{\dagger \, in } a_{i\oplus 1 }  
\nonumber \\
 & &-\left( a_{i\oplus 3}^\dagger a_{i\oplus 3 }^{in} 
 + a_{i\oplus 3 }^{\dagger \, in } a_{i\oplus 3 } \right)  
 -\left( a_{i\oplus 4 }^\dagger a_{i\oplus 4 }^{in} 
 + a_{i\oplus 4 }^{\dagger \, in } a_{i\oplus 4 } \right)  
				\label{G-}
\eeqa
When the input operators of the hexagonal modes are in the vacuum 
state, $\left\langle G_-(t) \right \rangle =0$. Moreover, by using 
\begin{description}
\item{i)} the commutation relations of input operators,
\item{ii)} the causality relations expressed by 
Eq. (\ref{causality}),
\item{iii)} the fact that the $a_i^{in}$ annihilates the vacuum on 
the right and $a_i^{\dagger \; in}$ 
the vacuum on the left,
\end{description}
it is not difficult to show that
\be
\left\langle G_-(t) G_-(t')   \right \rangle = \delta(t-t') 
\left\langle N_+ (t)  \right \rangle \; ,
							\label{G-corr}
\ee
with
\be
N_+ = N_i + N_{i \oplus 1} + N_{i \oplus 3}+N_{i \oplus 4} \; .
				\label{N+}   
\ee
At steady state $\left\langle N_+ (t)  \right \rangle $ 
does not depend on time and Eq.(\ref{dN-dt}) is
easily solved in the frequency domain. By setting 
\be
N_- (\omega) = \int \frac {d t }{\sqrt{2 \pi}} N_- (t) 
e^{i \omega t}\,, 
\ee
we have
\be
N_-(\omega) = \sqrt{2 \omega} 
\frac{G_- (\omega)}{2\gamma -i \omega}\,.
\ee 
By using the boundary relations for the outgoing mode operators 
$ a_i^{out }= a_i - \sqrt{2\gamma} a_i^{in}$,
we get
\be 
N_-^{out}(\omega) = 2\gamma N_- (\omega)  + N_-^{in}(\omega) -  
\sqrt{2\gamma} G_-(\omega)
= \sqrt{2\gamma} G_- (\omega) \frac{i \omega}{2\gamma - i \omega}
+ N_-^{in}(\omega) \,.
\ee
The fluctuation spectrum of the output photon 
number combination $N_-^{out} $ 
can now be calculated as:
\be
V_-(\omega) = \int d t \, 
\langle \delta N_-^{out}(t) N_-^{out} (0)\rangle e^{i \omega t}=
\gamma \langle N_+ \rangle \left(1 -\frac{ 4\gamma^2}{\omega^2 
+ 4 \gamma^2} \right) \; ,   							
									\label{Lorentzian} 
\ee
where $\delta N_-^{out}  =N_-^{out}  - \langle N_-^{out}  
\rangle=N_-^{out}$. Hence the noise 
spectrum of  $N_-^{out}$ turns out to have a standard 
Lorentzian shape 
\cite{LUCA92},
with the Lorentzian dip well below the shot-noise 
level represented by 
$2\gamma \langle N_+ \rangle = \langle N_+^{out} \rangle $. 
This result 
is reminiscent of the
\'' twin beam'' correlation spectrum calculated for the 
down-converted fields 
generated 
by a parametric oscillator without spatial aspects \cite{twin}. 
As a matter of fact,
in the case of twin beam generation, the photon number difference
between the two beams has zero commutator
with the system Hamiltonian. The calculation  of the noise spectrum 
of the number difference 
may be performed following exactly the same steps outlined here, 
and it is not a coincidence
that the result turns out to be the same. However, 
in contrast to the more standard method
of analysis, which
exploits the small quantum noise limit by linearizing the 
model equations around a classical steady-state,
the method used here is ``exact"\footnote{More precisely, 
it is exact as long as transverse modes other
than those forming the pattern at the instability onset  
are negligible.}, because it takes into account 
the full nonlinearity,
and does not depend on the particular steady state chosen.

In this connection it should be noted that our result 
concerning spatial intensity correlation
in the hexagonal pattern
is valid both above and below the threshold for pattern formation, 
where there is no classical pattern at all. Off-axis 
emission of photons is here
generated purely by quantum fluctuations; on average 
translational and rotational symmetry 
are preserved, so that off-axis photons can be emitted 
in any direction. 
However, the mechanism of photon absorption and emission 
is such to preserve the transverse radiation momentum; 
as a consequence any six regions in hexagonal arrangement 
with arbitrary orientation 
in the far field plane display exactly 
the same kind of intensity correlation that we have 
described above threshold.  

Incidentally, it is worth noting that the same method can 
be applied to analyze
the quantum spatial correlation in the far field plane of 
the down-converted field emitted by  
a optical parametric oscillator, 
where under proper circumstances the downconverted field is 
emitted in the form of two plane waves 
slightly tilted with respect to the 
cavity longitudinal axis~\cite{ROLLS}. By interference these 
two waves give rise
in the near field to a stripe pattern, while the far-field 
intensity distribution consists of two symmetrical spots.
From a microscopic point of view, the most relevant 
three-wave-mixing process 
close to threshold involves destruction of a pump photon 
that propagates longitudinally, 
and creation of two down-converted photons propagating 
in symmetrical directions, 
as it is required by transverse momentum conservation. 
Hence, the two off-axis signal waves 
emitted above threshold are made of twin 
photons and the photon numbers 
$N_1 , N_2$ that cross in the unit time two ideal 
detectors surroundings the
two spots in the far field plane are highly correlated 
\cite{LANG97,Marzoli97,NDOPO,JOPB99} . 
In the framework of a three-mode-model, valid close to 
threshold, a calculation analogous
to that performed in the case of hexagons shows that
the noise spectrum of their difference $N_- = N_2 - N_1$ 
is given exactly
by the formula (\ref{Lorentzian}). 

A last interesting remark is the following: 
in the Kerr cavity model, 
when the quantum fluctuation dynamics is linearized 
around the homogeneous steady-state below threshold, 
intensity correlation is found between any two
symmetric wave-vectors $\k$ and $-\k$ close to the critical circle,
just as in the case of parametric down-conversion. 
Hence the linear analysis
does not give any hint about the existence of a 
hexagonal pattern above threshold,
with its quantum correlations. 
This is not in contradiction with our result, since quantum
correlation between two symmetric modes obviously implies
a noise reduction in the observable $N_-$ connecting 
two couples of symmetric modes.
However, the inverse is not in general true.
Since the intensity difference between symmetric modes is not 
a constant of motion of the full Hamiltonian, we expect 
that the level of correlation between symmetric modes 
decreases when approaching the region where 
linearization fails, that is,
in the neighbourhood of the instability point, 
where the size of fluctuations
increases, or for a truly microscopic system 
characterized by a small
saturation photon number parameter. On the contrary, our
non-linear analysis shows that the correlations 
which are at the origin
of noise suppression in $N_-$ do not depend on 
the distance from theshold or
on the system size.

\section{Phase Quadrature correlation in the hexagonal pattern}
\label{phquad} 

In the previous section it has been demonstrated a 
high level of intensity correlation
in groups of four among the hexagonal modes. A natural question 
which arises is whether
there exist any other kind of correlation among the six modes 
at the quantum level.
This possibility is suggested by the example of twin beams 
generation. In fact, in this case,
not only the photon number fluctuations are correlated, 
but also the ``phase" fluctuations
of the two beams are anti-correlated at a quantum level. 
To be more precise, there 
exist two orthogonal field quadratures of the two beams 
which show, at the same time,
a high level of (anti)correlation. This is at the origin 
of the EPR aspects of the twin beams \cite{EPR,MYEPR},
which are widely exploited in the field of quantum information
 with continuous variables (see e.g.\cite{TELE,CRYPTO,SETH}). 

Since we do not have in mind other constant of motion 
(other observable commuting with the Hamiltonian), 
we have to resort
to traditional means of calculations. Namely, in the framework 
of the seven-mode discrete model (\ref{expansion}),
we separate  
the  quantum fluctuation
operator from the classical mean field
\be {}
a_i(t)= \frac{1}{\sqrt{g}} \beta_i  + \delta a_i( t) 
\qquad i=0,1, \ldots 6
			 \label{small}
\ee
where $\beta_i$ are the classical steady state amplitudes of 
the seven mode. More precisely,
they are the steady state solution of a set of  classical 
dynamical equation, obtained
by introducing in Eq.(\ref{eqLL}) the discrete expansion of 
the envelope operator
$E(\x,t) = \frac{1}{b} \alpha_0 (t) + \frac{1}{b} 
\sum_{i=1}^{6} \alpha_i (t) e^{i \k_i \cdot \x}$.
These equations have the form:
\begin{eqnarray}
\frac{1}{\gamma}\frac{d}{dt} \alpha_0 &=& E_I-(1+i\Delta)\alpha_0
+i\alpha^*_0a_0^2
+2i\alpha _0\sum_{j=1}^{6} \alpha _j^*\alpha_j
+i \alpha_0^*\sum_{j=1}^{6}\alpha_j  \alpha_{j\oplus 3}
+2i \sum_{j=1}^{6}\alpha^*_j 
\alpha_{j\oplus 1} \alpha_{j\oplus 5}\, ,
							\nonumber\\
\frac{1}{\gamma} \frac{d}{dt} \alpha_j&=&-(1+2i)\alpha_j
+i \alpha^*_j \alpha^2_j
+2i \alpha_j|\alpha_0|^2
+2i \alpha_j \sum_{i\ne j=1}^{6} \alpha^*_i \alpha_i
+2i \alpha^*_{j\oplus 3}\left(
\alpha_{j\oplus 4}\alpha_{j\oplus 1}+ \alpha_{j\oplus 5} 
\alpha_{j\oplus 2}\right)
\nonumber\\
&+& i\alpha_0^2 \alpha^*_{j\oplus 3}
+2i\alpha^*_0 \alpha_{j\oplus 5} \alpha_{j\oplus 1}
+2i\alpha_0\left(\alpha^*_{j\oplus 4} \alpha_{j\oplus 5}
+ \alpha_{j\oplus 1} \alpha^*_{j\oplus 2}\right)
\,,\qquad j=1,\ldots,6\; ,
						\label{cdyn}
\end{eqnarray}
where we took into account the fact that all 
the hexagonal modes have 
the same critical transverse wave number, such that 
$\Delta + l_D^2k_c^2=2$.

By inserting the expansion (\ref{expansion}) into the 
model equation  (\ref{Aeq}), with
the ansatz  (\ref{small}), and keeping
only the leading terms in the small quantum fluctuations 
we are left with a problem of the form
\be   
\frac{d}{dt} 
\left( \begin{array} {c} \delta a_0\\ \vdots \\  \delta a_6 \\   
				\delta a_0^\dagger\\ \vdots \\ \delta a_6^\dagger 
	\end{array} \right)
=
\left( \begin{array}{c}  \\14 \times 14 \\ \mbox{matrix}  \\   \\ 
\mbox{matrix elements}  \\   
				 \mbox{are functions of } \\   
				 (\beta_0,\beta_0^* \ldots, \beta_6,\beta_6^*) \\
   \end{array}  \right)  \;
\left( \begin{array} {c} \delta a_0\\ \vdots \\  \delta a_6 \\   
				\delta a_0^\dagger\\ \vdots \\ \delta a_6^\dagger 
	\end{array} \right) +
\sqrt{2\gamma}
\left( \begin{array} {c} \delta a_0^{in} \\ 
\vdots \\  \delta a_6^{in}  \\   
				\delta a_0^{\dagger\; in} \\ 
				\vdots \\ \delta a_6^{\dagger\; in} 
	\end{array} \right)\; ,
						\label{linear}
\ee
where the explicit form of the matrix elements in terms of 
the classical steady state
amplitudes $(\beta_0,\beta_0^* \ldots, \beta_6,\beta_6^*)$ 
is given in Appendix A.
The input fluctuation operators at RHS are in the vacuum state 
and have commutation relations 
as those in Eq.(\ref{incomm}).
This represents a $14 \times 14$ linear problem, which is trivial 
from a numeric point of view; 
however, when searching for some explicit combination of mode 
operators that has sub-shot-noise
fluctuations, the problem seems too complex 
to find analytical solutions. 
Nevertheless, some hints came from the analysis of 
the classical steady state. 

\begin{figure}[t]
\centerline{\epsfig{figure=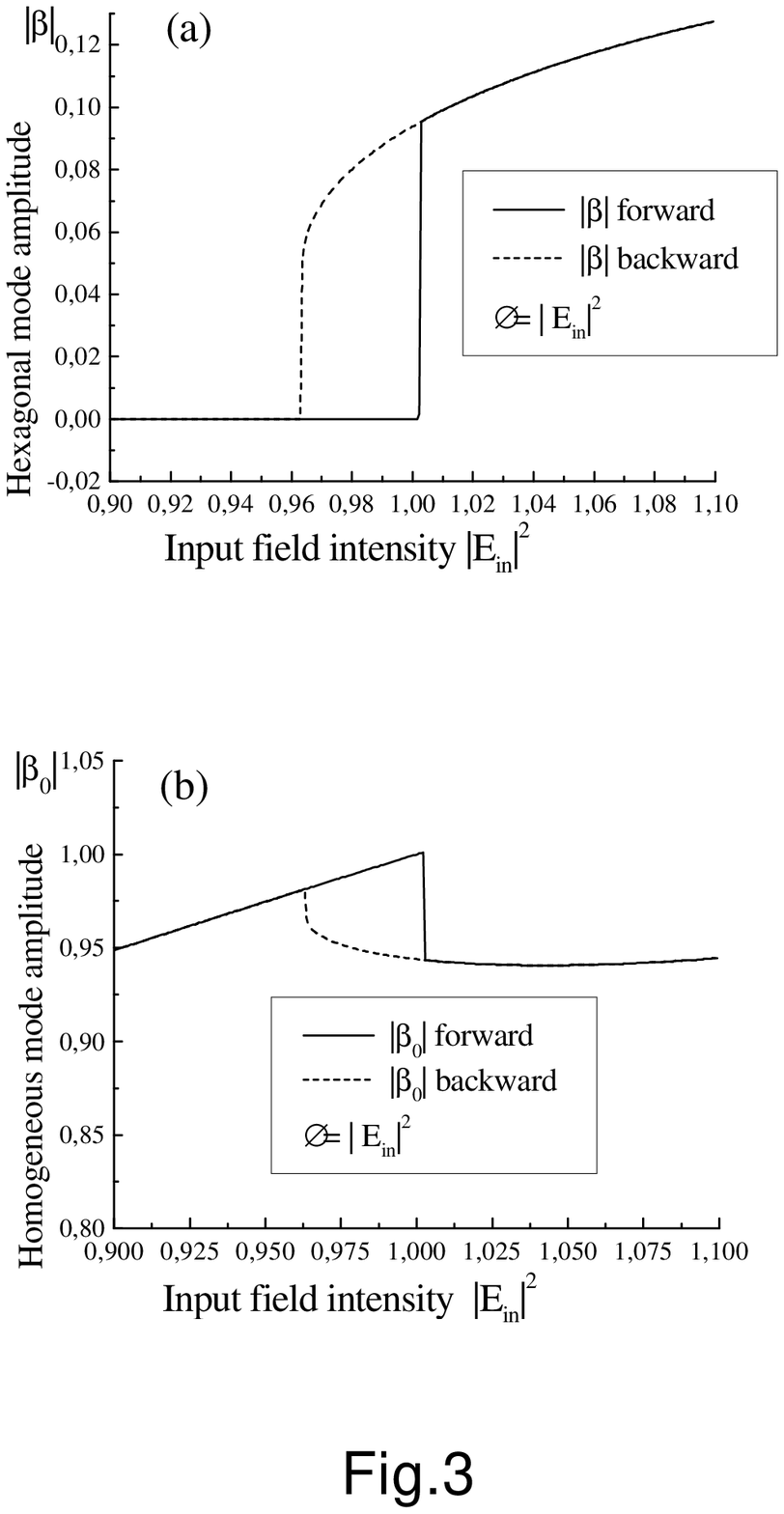,width=14cm}}
\caption{Example of the steady-state curves 
for the hexagonal mode (a)
and the homogeneous mode(b) amplitudes as
a function of the input beam intensity. Solid (dashed) lines 
are obtained by 
a numerical integration of Eqs.(\ref{cdyn}) over times long
compared to time scales of dynamical evolution,   
and  by a slow forward (backward) sweep of the input beam intensity
across the instability threshold ($|E_{in}|^2=1$). 
Cavity detuning is chosen as $\Delta= |E_{0s}|^2$.} 
\label{fig3}
\end{figure}

We integrated numerically the set of classical equations 
(\ref{cdyn}), and looked at the long time behaviour
of the mode amplitudes. 
Above the critical point (\ref{critical})
the hexagonal steady-state of the classical equations of the model
\be {}
E_s(\x)=  \frac{1}{b} \beta_0  + \frac{1}{b} \sum_{i=1}^{6} \beta_i  
e^{i \k_i \cdot \x}  \; ,
					\label{ssexpansion}
\ee
is characterized by:
\begin{description}
\item{--} 
All the hexagonal modes have the same mean intensity
\be 
\left|\beta_1\right|=\left|\beta_2\right|
=\ldots \left|\beta_6\right|:= 
\left|\beta\right|\,,
						\label{intensity}
\ee
so that we set $\beta_j= \left|\beta \right| 
e^{i \varphi_j}$. Figure \ref{fig3}a 
shows the steady-state  hexagonal mode amplitude $|\beta|$ as a 
function of the input field intensity.
The figure is obtained by integrating the classical model equations 
(\ref{cdyn}) under 
a slow (with respect to the characteristic time needed to reach 
the steady state)
increasing of the input field intensity across the instability 
threshold (solid curve);
the dashed curve correspond  to a slow decrease of the input 
intensity. As it is well know \cite{FIRTH92,SCRO94} 
the instability is subcritical, and the hexagonal mode amplitude 
shows the typical hysteresis cycle.
Figure \ref{fig3}b is the same for the amplitude $|\beta_0|$ of 
the homogeneous mode.
\item{--} 
The sums of the phases of symmetric modes are all equal
\be
\varphi_1+\varphi_4=\varphi_3+\varphi_6=\varphi_5+\varphi_2: 
= 2 \phi\,. 
						\label{sumphases}
\ee
\item{--} 
The differences of the phases of symmetric modes 
sum up to zero\footnote{Notice that in the classical
 analysis presented in \cite{FIRTH92,SCRO94}, some initial 
 assumptions
on the steady state 
were made so that condition (\ref{sumphases})
 was automatically fulfilled,with $\phi=0$, and condition 
 (\ref{diffphases}) was written
as $(\Delta \varphi_1 + \Delta \varphi_3 + 
\Delta \varphi_5)/2=0,\pi$.}
\be
\Delta \varphi_1 + \Delta \varphi_3 + 
\Delta \varphi_5 = 0 \quad \mbox{with} \quad \Delta \varphi_j = 
\varphi_j-\varphi_{j\oplus 3}\,.
					\label{diffphases}
\ee
\end{description}
There are two phases which are not fixed by the 
steady-state equations, namely
two of the phase differences between symmetric modes, say 
$\Delta \varphi_1$ and $\Delta \varphi_3$.
The value of these phases at steady-state depends only on 
initial conditions, and these variables are
dominated by noise. This circumstance is  a noteworthy 
consequence of the translational symmetry of the model, broken
by the formation of the pattern in the transverse plane. 
As a matter of fact, for each value of
the input field intensity $|E_I|^2$ above the critical point, 
there exists a continuous set 
of steady-state solutions of the model equations (\ref{cdyn}), 
of the form
\beqa {  }
E_s(\x)
&=&  \frac{1}{b} \beta_0  + \frac{2}{b}  |\beta| e^{i \phi}
      \left[ 
     \cos {\left( \k_1 \cdot \x  + \Delta \varphi_1 \right)}
      +\cos {\left( \k_3 \cdot \x  + \Delta \varphi_3 \right)} +
       \cos {\left(\k_5 \cdot \x  
       - (\Delta \varphi_1 +\Delta \varphi_3) \right)}
      \right]
					\label{set1}
\eeqa
corresponding to an arbitrary choice of $\Delta \varphi_1$, 
$\Delta \varphi_3$. By setting
$\Delta \varphi_1=\k_1 \cdot \vec{\Delta x}$, 
and $\Delta \varphi_3=\k_3 \cdot \vec{\Delta x}$,
we notice that  $\Delta \varphi_1 +\Delta \varphi_3 
= (\k_1+\k_3) \cdot \vec{\Delta x}= -\k_5 \cdot\vec{\Delta x}$.
Thus the set of solutions (\ref{set1}) corresponds to 
the  continuous set of rigid translations of the hexagonal
pattern in the $(x,y)$ plane,  $\x \to \x + \vec{\Delta x}$.\\
As it is common in continuos symmetry breaking phase transition 
\cite{FORSTER} close to the critical point, the noise
is concentrated on the mode that aims to restore the symmetry 
broken by the transition. In our
case we argue that close to the instability threshold, 
quantum fluctuations corresponding to rigid
translations of the pattern result in huge fluctuations 
of the differences of the phases between symmetric
modes $\Delta \varphi_j$, but leaves invariant the quantities
\beqa
&(A)& \quad \Delta \varphi_1 +\Delta \varphi_3 
+\Delta \varphi_5   \label{A}\\
&(B)& \quad (\varphi_i +\varphi_{i\oplus 3})  
- (\varphi_{i\oplus 1}   +\varphi_{i\oplus 4})  
\quad i=1, \ldots 6
					\label{B}
\eeqa 
Hence, our possible candidates for squeezing are the observables
\be
 \left( \delta a_1 -\delta a_4 + \delta a_3 -\delta a_6 
 +\delta a_5 -\delta a_2 \right) \rm e^{ -i \theta}
+{\rm h.c.} \; ,
			\label{AA}
\ee 
which corresponds to the classical quantity (A) in (\ref{A})
for $\theta=\phi +\pi/2$, and
\be
\left[ \left(\delta a_i +\delta a_{i \oplus 3} \right)
-  \left(\delta a_{i \oplus 1}  +\delta a_{i \oplus 4} \right)
   \right] \rm e^{-i \psi } + {\rm h.c.}   \; , 
			\label{BB}
\ee 
which corresponds to (B) in (\ref{B}) when 
$\psi=\phi +\pi/2$. 
In these definitions $\theta, \psi$ are left
arbitrary and will be used as optimization parameters. 

In addition, an obvious observable to take into account is
\be
\left[ \left(\delta a_i - \delta a_{i \oplus 3} \right)
+ \left(\delta a_{i \oplus 1}  -\delta a_{i \oplus 4} \right)
   \right] \rm e^{ i \phi } -{\rm h.c.}  \quad i=1,2 \; , 
										\label{CC}
\ee 
In the small quantum noise approximation, this observable 
corresponds to the the photon number difference 
$N_-$, which was shown in the previous section to have a 
sub-shot noise fluctuation spectrum.

Luckily enough, when considering the combinations of mode 
quadratures given by (\ref{AA}), (\ref{BB}), (\ref{CC}), 
their dynamical equations decouple from the other equations 
of the system (\ref{linear}),
and we are left with $2\times 2$ linear problems. 

Before examining specific cases, we note that 
the quantum dynamics of a generic linear system can be written as
\begin{equation} 
\frac{d}{dt} {\cal V}={\bf M} {\cal V}
+\sqrt{2 \gamma}{\cal V}^{\rm in}\,,
					\label{linsys}
\end{equation}
where ${\cal V}$, ${\cal V}^{\rm in}$ are the system 
and noise operators vectors   
respectively,
while ${\bf M}$ is a coefficient matrix.

Suppose that the
components of the vector ${\cal V}$  are two 
conjugate field quadratures $Z(0)$ and $Z(\pi/2)$; 
at steady state the system 
can be solved in the frequency domain, and, with 
the aid of an input-output relation \cite{GAR}
\begin{equation} 
{\cal V}^{\rm out}=\sqrt{2\gamma} {\cal V}- {\cal V}^{\rm in}\,,
					\label{inout}
\end{equation}
it is possible to calculate the output correlation matrix.
In the frequency domain it reads
\begin{equation}\label{Cout}
{\bf C}^{\rm out}(\omega)
 \equiv \int \rm d \omega' \langle {\cal V}^{\rm out}(\omega) 
 [{\cal V}^{\rm out}]^T (\omega')  \rangle
=\left[2 \gamma\left({\bf M}+i\omega{\bf I} \right)^{-1} 
+ {\bf I}\right]
{\bf C}^{\rm in} (\omega)
\left[2\gamma \left({\bf M}-i\omega{\bf I}\right)^{-1}
+{\bf I}\right]^T\,,
\end{equation}
where ${\bf I}$ is the identity matrix, 
$T$ means the transpose, and
${\bf C}^{\rm in} (\omega) =\int \rm d \omega' 
\langle {\cal V}^{\rm in}(\omega)\,
[{\cal V}^{\rm in}]^T (\omega')\rangle$ is the
input correlation matrix.

Moreover, the noise spectrum of  a generic quadrature 
\begin{equation}
Z({\psi }) =Z(0)\, \cos\psi+Z(\pi/2)\, \sin\psi\,,
\end{equation}
is given by 
\begin{eqnarray}
{\cal S}_Z (\psi,\omega) 
&:=& \int \rm d \omega' 
\langle {\delta Z}^{\rm out}(\psi,\omega)
\,{\delta Z}^{\rm out}(\psi,\omega') \rangle \\
&=&
{\bf C}^{\rm out}_{1,1}(\omega)\cos^2\psi
+{\bf C}^{\rm out}_{2,2}(\omega)\sin^2\psi
+\left({\bf C}^{\rm out}_{1,2}(\omega)
+{\bf C}^{\rm out}_{2,1}(\omega)\right)\sin\psi
\cos\psi\,.
\end{eqnarray}

With the above in mind we are now going to consider specific cases.

\subsection{Noise in the sum of phase differences}
\label{secA}

Let us consider the quadrature operator
\begin{equation}
\label {Wtheta}
W(\theta)=\frac {1} {\sqrt{6}}  
\left(  a_1 - a_4 + a_3 -a_6 + a_5 - a_2 \right) e^{ -\im \theta}
+{\rm h.c.} \; ,
\end{equation}
The dynamics of the fluctuation vector ${\cal V  } 
=[\delta W(0),\delta W(\pi/2)]^T$,
gives rise to a closed system like (\ref{linsys}),
with 
\begin{eqnarray}\label{MCpm}
{\bf M}=\gamma 
\left[
\begin{array}{cc}
{\rm Re}\left\{{\cal A}_+\right\} &
-{\rm Im} \left\{{\cal A}_+\right\}
\\
{\rm Im}\left\{{\cal A}_-\right\}&
{\rm Re} \left\{{\cal A}_-\right\}
\end{array}
\right]
\,,
\end{eqnarray}
and
\begin{eqnarray} 
{\cal A}_{\pm}&=&\left[-1-2i+2i|\beta_0|^2+10i|\beta|^2
-4i\beta_0\beta^*-4i\beta_0^*\beta\right]
\nonumber\\
&\pm&\left[-4i\beta_0^*\beta^*+5i(\beta^*)^2
+i(\beta_0^*)^2\right]\,.
						\label{Cpm}
\end{eqnarray}
In writing (\ref{Cpm}) the symmetries of the steady state have been 
taken into account; moreover,
among the possible stationary states (\ref{set1}) we used the one 
corresponding to
$\Delta \varphi_1=0$, $\Delta \varphi_3=0$. 
The input correlation matrix is given by
\begin{equation}
\label{Cin}
{\bf C}^{\rm in}
=\left[
\begin{array}{cc}
1 & i \\
-i& 1 
\end{array}
\right]
\,.
\end{equation}

We performed calculations of the noise spectrum 
${\cal  S}_W(\theta,\omega)$   
for various values of the input beam intensity $|E_{in}|^2$ in the 
region where the hexagonal solution
exists. 
The amplitudes of the homogeneous mode  $\beta_0$, and of the 
hexagonal mode $\beta$ were obtained
both by numerically integrating the dynamical classical equations 
(\ref{cdyn}) and looking at the long
time behavior, and by numerically
solving the nonlinear steady-state equations (see Appendix B). 
An excellent agreement between the two approaches
was found.

\begin{figure}[t]
\centerline{\epsfig{figure=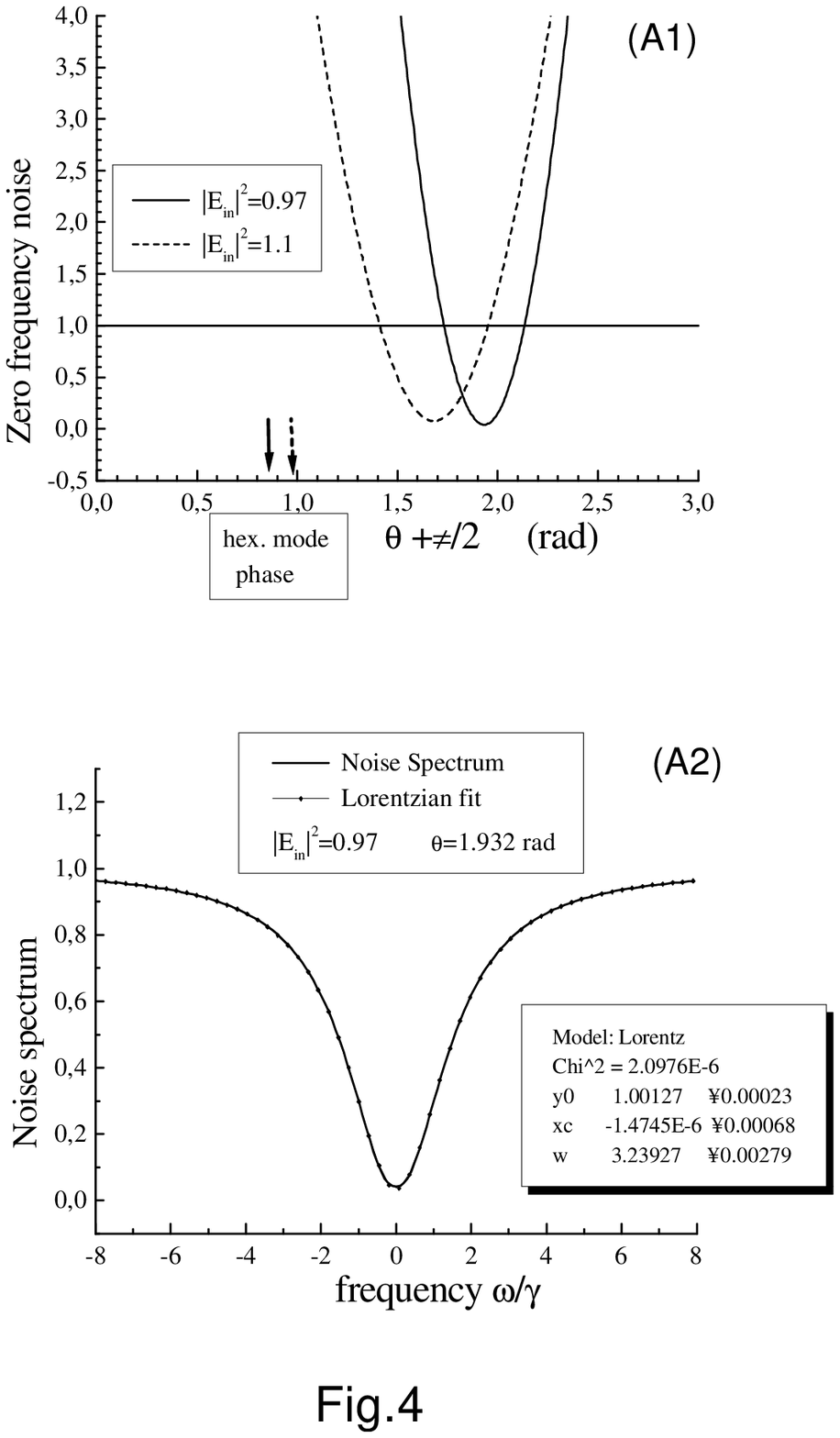,width=14cm}}
\caption{
Part (A1) shows the zero frequency spectrum of the observable
$W(\theta)$, given by Eq.(\ref{Wtheta}),
vs the quadrature angle $\theta$ shifted by $\pi/2$, for 
two values of the input field.
The arrows indicate the corresponding hexagonal mode phases.
Part (A2) shows the frequency spectrum for 
the optimal value of quadrature angle.
$\Delta= |E_{0s}|^2$. Other parameters are indicated 
in the figure. Dots correspond to a Lorentzian fit.}
\label{fig4}
\end{figure}

Figure \ref{fig4} shows an example of the typical results. 
Part A1 of the figure plots the zero frequency spectrum 
as a function of the quadrature angle $\theta$, shifted by $\pi/2$. 
Sub shot-noise fluctuations are present
for a rather broad range of quadrature angles; it has to be noticed  
that the quadrature operator exactly
corresponding to
the classical sum of phase differences (\ref{A}) is the one with 
$\theta +\pi/2=\phi$ where $\phi$
is the steady-state phase of $\beta$ 
(indicated by arrows in the figure). 
Hence sub shot noise fluctuations
are present for quadratures somehow rotated with respect to the 
semiclassical "phase" quadrature. In  Part
A2 of the figure the quadrature angle 
is chosen to optimize squeezing; 
the plots evidenciate the typical
Lorentzian shape of the fluctuation spectrum.

Line (A) in figure \ref{fig7} plots the best squeezing (that is, 
the low frequency noise with the phase $\theta$ optimized)
as a function of the input field intensity, and shows that 
sub shot noise fluctuations
for this observable are present in the whole region where the 
hexagonal solution exists.

\subsection{Noise in the difference of phase sums}
\label{secB}

We consider now the  quadrature operators
\begin{equation}\label{Qpsi}
Q (\psi)= \frac {1} {\sqrt{4}}  \left(a_i + a_{i \oplus 3}
-a_{i \oplus 1}-a_{i \oplus 4}
\right)e^{-i\psi}+{\rm h.c.}\,, 
\end{equation}
In fact there are two independent mode combinations of this 
form, e.g. for $i=1,2$, and for both
of them
we get a closed system like (\ref{linsys})
for the fluctuation vector ${\cal V}
=[\delta Q(0),\delta Q(\pi/2)]^T$, with
\begin{equation}\label{MApm}
{\bf M}=\gamma
\left[
\begin{array}{cc}
{\rm Re}\left\{{\cal B}_+\right\}&
-{\rm Im} \left\{{\cal B}_+\right\}
\\
{\rm Im}\left\{{\cal B}_-\right\}&
{\rm Re} \left\{{\cal B}_-\right\}
\end{array}
\right]\,,
\end{equation}
and
\begin{eqnarray}\label{Apm}
{\cal B}_{\pm}&=&\left[-1-2i+2i|\beta_0|^2+6i|\beta|^2
-2i\beta_0\beta^*-2i\beta_0^*\beta\right]
\nonumber\\
&\pm&\left[2i\beta_0^*\beta^*-3i(\beta^*)^2-i(\beta_0^*)^2\right]\,.
\end{eqnarray}

\begin{figure}[t]
\centerline{\epsfig{figure=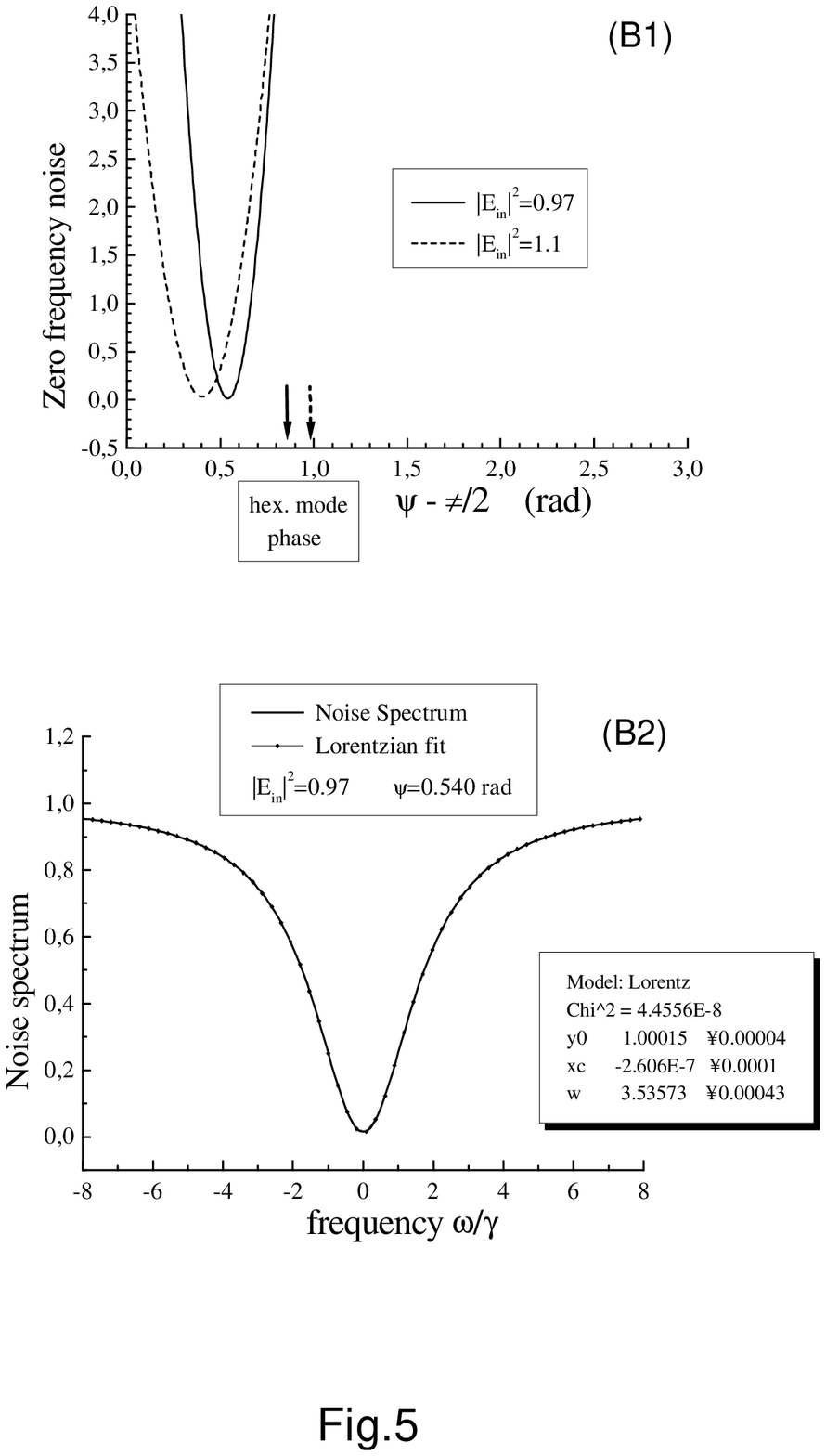,width=14cm}}
\caption[]{
Part (B1) shows the zero frequency spectrum of 
the observable $Q(\psi)$ given
by Eq.(\ref{Qpsi}),
vs the quadrature angle $\psi$ shifted by $\pi/2$, for 
two values of the input field.
The arrows indicate the corresponding hexagonal mode phases. 
Part (B2) shows the frequency spectrum for 
the optimal value of quadrature angle.$\Delta= |E_{0s}|^2$. 
Other parameters are indicated 
in the figure. Dots correspond to a Lorentzian fit.}
\label{fig5}
\end{figure}

The input noise correlations are again given by (\ref{Cin}). 
Fig.\ref{fig5} shows the  noise
spectrum ${\cal S}_Q(\psi,\omega)$, 
and is analogous to  Fig.\ref{fig4}. 
As in the previous case, the spectrum has a 
Lorentzian shape and exhibits squeezing
in the whole region where
hexagons are predicted by the seven mode model, 
although the squeezed quadratures do not
exactly coincide with the one corresponding to 
the classical phases in (\ref{B}).

Line (B) in figure \ref{fig7} plots the best squeezing 
as a function of the input field intensity, 
and shows that also in this case
sub shot noise fluctuations
for this observable are present in the whole region where the 
hexagonal solution exists.

\subsection{Noise in the sum of intensity differences}
\label{secC}

Finally, let us consider the two independent quadratures
\begin{equation}\label{Xpsi}
X(\psi)=
\frac {1} {\sqrt{4}}  \left[ \left(a_i +a_{i \oplus 3} \right)
-  \left(a_{i \oplus 1}  +a_{i \oplus 4} \right)
   \right] e^{-\im \psi } + {\rm h.c.}  \quad i=1,2 \; , 
\end{equation}
Again, their fluctuation vectors  
${\cal V}=[\delta X(0),\delta X(\pi/2)]^T$,
give rise to two closed systems like (\ref{linsys})
with 
\begin{eqnarray}\label{MBpm}
{\bf M}=\gamma
\left[
\begin{array}{cc}
{\rm Re}\left\{{\cal C}_+\right\}&
-{\rm Im} \left\{{\cal C}_+\right\}
\\
{\rm Im}\left\{{\cal C}_-\right\}&
{\rm Re} \left\{{\cal C}_-\right\}
\end{array}
\right]\,,
\end{eqnarray}
and
\begin{eqnarray}  \label{Bpm}
{\cal C}_{\pm}&=&\left[-1-2i+2i|\beta_0|^2+10i|\beta|^2
+2i\beta_0\beta^*+2i\beta_0^*\beta\right]
\nonumber\\
&\pm&\left[2i\beta_0^*\beta^*+5i(\beta^*)^2
+i(\beta_0^*)^2\right]\,.
\end{eqnarray}

The input correlation matrix is the same of Eq.(\ref{Cin}).

It is worth noting that the quadrature corresponding 
to the hexagonal mode stationary phase $\phi$
satisfies a simple equation
\begin{equation}
\frac{d}{dt}\,\delta X(\phi)=-2\gamma\,\delta X(\phi)
+\sqrt{2\gamma}\,\delta X^{\rm in}\,,
\end{equation}
which gives a Lorentzian shape spectrum
\begin{equation}
{\cal S}_X(\phi,\omega) 
=\frac{\omega^2}{4\gamma^2+\omega^2}\,.
			\label{Lor2}
\end{equation}
Taking into account that now the shot-noise level is scaled to 1,
this is in perfect agreement with the results of Section II, since
$\delta X(\phi)$ represents the linearized version of the 
observable $N_-$. 

\begin{figure}[t]
\centerline{\epsfig{figure=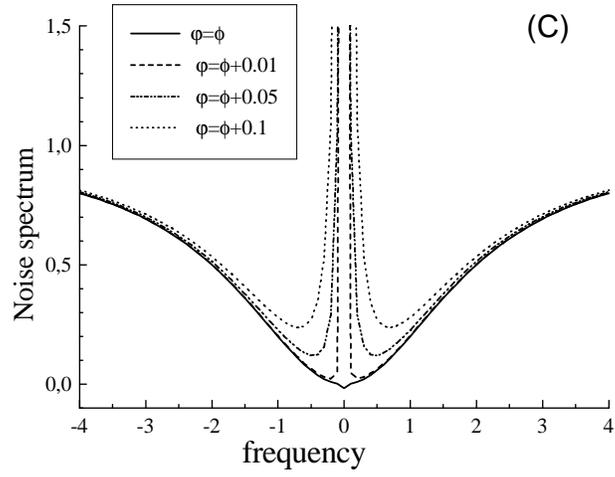,width=14cm}}
\caption{Noise spectrum of the observable $X(\varphi)$ 
given by Eq.(\ref{Xpsi}). 
The solid line is the Lorentzian spectrum when the 
quadrature angle  $\varphi=\phi$; 
the other lines correspond to quadratures slightly 
rotated with respect to the
amplitude quadrature, as indicated in the figure.}
\label{fig6}
\end{figure}

\begin{figure}[t]
\centerline{\epsfig{figure=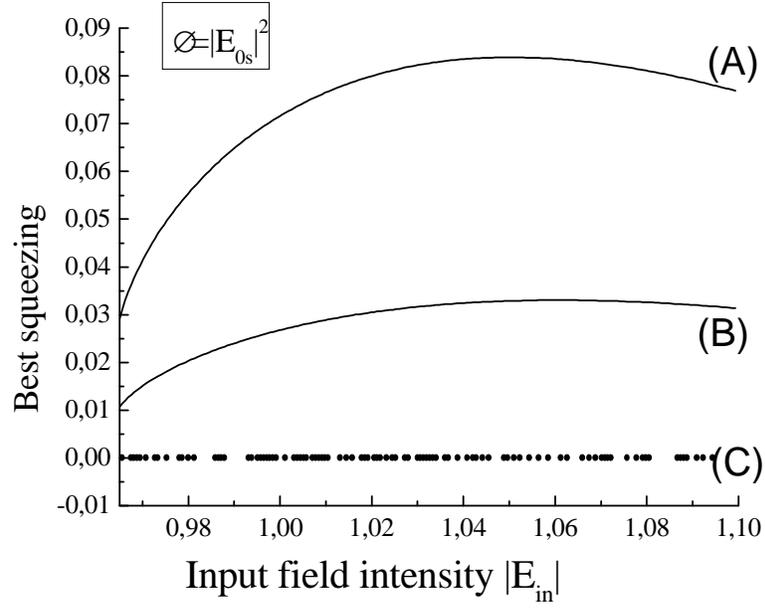,width=14cm}}
\caption{Best squeezing (low frequency noise with 
the quadrature angle optimized) 
as a function of the input field intensity, 
for the three observables considered in Sections 
\ref{secA} (line A),
\ref{secB} (line B), and \ref{secC}  (line C). }
\label{fig7}
\end{figure}

It is interesting to remark that, differently
with respect to the previous cases, for this 
combination of modes the low frequency noise 
reduction abruptly disappears
when the quadrature is slightly rotated with 
respect to the amplitude quadrature. In fact,
as shown by figure \ref{fig6}, a large positive 
peak located at zero frequency 
appears on the Lorentzian spectrum when the 
quadrature angle is shifted from 
$\phi$. 

Line (C) in figure \ref{fig7} plots a numerical evaluation of 
the best squeezing 
as a function of the input field intensity. As it is obvious from
the analytical formula \ref{Lor2}) complete 
noise suppression is predicted 
at zero frequency for $\varphi=\phi$ (the amplitude quadrature),
in the whole region where the 
hexagonal solution exists.

\section{Concluding remarks and open questions}
\label{conclusions} 

In conclusion, we have studied the quantum features underlying the 
hexagonal pattern formation in a Kerr cavity. 
We have identified a rich scenario of purely 
quantum correlations among the 
off axis modes that form the hexagonal pattern. In fact 
we have shown that at least five independent 
combinations of mode observables
exhibit sub shot-noise fluctuations. 

We believe that the six-mode entanglement analysed in this paper 
might be of great interest. On the one side  this kind of patterns
are widely studied in labs, so that 
it should be easy to look for the 
described quantum effects. On the other side, the quantum properties 
of such systems make them  good candidates for quantum information 
processing with continuous variable \cite{TELE,CRYPTO,SETH,SAM}.
In particular, the existence of several observables showing 
correlations  at the quantum level opens the 
possibility of distilling a considerable amount of entanglement 
in practical situations, and to use it.

In this connection an interesting question is whether this kind
of spatial entanglement is characteristic of the particular
model analysed here, or it is in general 
typical of the hexagonal pattern,
independently of the detailed mechanism that underlie its formation.
Further investigations about models 
closer to experimental situations where hexagons heve been found 
are hence in order.

An other interesting point
concerns the possibility that some squeezing 
is missing from our picture.
The symmetry of the problem suggests the possibility that  
quantum correlations - more probably anti-correlations-exist
between the the sum of all hexagonal modes 
and the pump mode. This would be closely similar to
the situation analysed in \cite{SINATRA,ZAMBRINI}, in the context
of a self-defocusing Kerr cavity, 
where quantum anticorrelations were found
between the pump on-axis intensity 
and the intensities of two symmetrical off-axis
modes. 

A final open question is whether these quantum correlations
survive when including in the model the full 
continuum of transverse modes.
As a matter of fact, being the bifurcation 
subcritical, the hexagonal mode
emerges at threshold with a finite amplitude. Hence 
higher order spatial harmonics might have a 
non-negligible amplitude evening in the neighbourhood
of the threshold, and photon scattering in 
these additional modes might spoil
our correlations. This will be the subject 
of future numerical investigations.

{\acknowledgements}

The authors warmly thank
L.A. Lugiato for useful suggestions and costant encouragement. 
This work was carried out in the framework 
of the network QSTRUCT of the 
EU TMR programme.
S.M. acknoledges financial
support from the MURST project 
\''Spatial pattern control in nonlinear optical systems"

\section*{Appendix A}

In this Appendix we give the explicit form of 
the linear system (\ref{linear})
for the dynamics of fluctuations of the hexagonal modes. 
In writing these equations we
took into account the symmetries of the hexagonal 
steady state expressed by conditions
(\ref{intensity}),(\ref{sumphases}) and 
(\ref{diffphases}); moreover among the possible 
steady states in (\ref{set1}) we choose for simplicity the
one with $\Delta \varphi_j=0$ , $j=1,3,5$ 
(this correspond to fixing the origin of the coordinate axes
in the transverse plane). We get
\begin{eqnarray}\label{linqLeqs}
\frac{1}{\gamma} \frac{d}{dt}\delta a_0&=&
\left[-1-i\theta+2i|\beta_0|^2+12i|\beta|^2\right]\delta a_0
+\left[i\beta_0^2+6i\beta^2\right]\delta a_0^{\dag}
\nonumber\\
&+&\left[2i\beta_0\beta^*+2i\beta_0^*\beta+4i|\beta|^2\right]
\sum_{j=1}^6\delta a_j
+\left[2i\beta_0\beta\right]\sum_{j=1}^6\delta a_j^{\dag}  
+\sqrt{ \frac{2}{\gamma}}\,\delta a_0^{\rm in}\,,
\\
\frac{1}{\gamma} \frac{d}{dt}\delta a_j&=&
\left[2i\beta_0\beta^*+2i\beta_0^*\beta+4i|\beta|^2\right]\delta a_0
+\left[2i\beta_0\beta+2i\beta^2\right]\delta a_0^{\dag}
\nonumber\\
&+&\left[-1-2i+2i|\beta_0|^2+12i|\beta|^2\right] \delta a_j
+\left[i\beta^2\right] \delta a_j^{\dag}
\nonumber\\
&+&\left[2i\beta_0\beta^*+2i\beta_O^*\beta+4i|\beta|^2\right] 
\delta a_{j\oplus 1}
+\left[2i\beta^2\right] \delta a_{j\oplus 1}^{\dag}
\nonumber\\
&+&\left[4i|\beta|^2\right] \delta a_{j\oplus 2}
+\left[2i \beta_0\beta+2i\beta^2\right] \delta a_{j\oplus 2}^{\dag}
\nonumber\\
&+&\left[2i|\beta|^2\right] \delta a_{j\oplus 3}
+\left[i \beta_0^2+6i\beta^2\right] \delta a_{j\oplus 3}^{\dag}
\nonumber\\
&+&\left[4i|\beta|^2\right] \delta a_{j\oplus 4}
+\left[2i\beta_0\beta+2i\beta^2\right] \delta a_{j\oplus 4}^{\dag}
\nonumber\\
&+&\left[2i\beta_0\beta^*+2i\beta_0^*\beta+4i|\beta|^2\right] 
\delta a_{j\oplus 5}
+\left[2i\beta^2\right] \delta a_{j \oplus 5}^{\dag}
+\sqrt{ \frac{2}{\gamma}}\,\delta a_j^{\rm in}\,,
\qquad j=1,\ldots,6\,.
\end{eqnarray}

The elements of the matrices ${\bf M}$ can be
built up by appropriately combining some of the above equations.

\section*{Appendix B}

This appendix gives the explicit expression of 
the nonlinear equations
for the classical hexagonal steady-state. 
These are obtained by equating to zero the
l.h.s of Eqs.(\ref{cdyn}); in order to find a numerical solution 
we transformed them in four nonlinear equations 
for the real variables $\{u_0,v_0,u_1,v_1\}$
introduced as \cite{SCRO94}
\begin{eqnarray}\label{uv}
\beta_0&=&E_{0s}+E_{0s}\left(u_0+iv_0\right)\,,
\\
\beta&=&\frac{E_{0s}}{2\sqrt{3}}\left(u_1+iv_1\right)\,.
\end{eqnarray}
The resulting equations are
\begin{eqnarray}\label{u0v0u1v1}
0&=&-u_0+(\Delta-|E_{0s}|^2)v_0-|E_{0s}|^2\Bigg(
2u_0v_0+u_1v_1+u_0^2v_0+\frac{1}{2}v_0u_1^2
\nonumber\\
&+&\frac{1}{2\sqrt{3}}u_1^2v_1+u_0u_1v_1+v_0^3+\frac{3}{2}v_0v_1^2
+\frac{1}{2\sqrt{3}}v_1^3\Bigg)\,,
\\
0&=&-v_0-(\Delta-|E_{0s}|^2)u_0+|E_{0s}|^2\Bigg(
2u_0+3u_0^2+\frac{3}{2}u_1^2+v_0^2+\frac{1}{2}v_1^2+u_0^3
\nonumber\\
&+&\frac{3}{2}u_0u_1^2\frac{1}{2\sqrt{3}}u_1^3+u_0v_0^2
+\frac{1}{2}u_0v_1^2
+v_0u_1v_1+\frac{1}{2\sqrt{3}}u_1v_1^2\Bigg)\,,
\\
0&=&-u_1+2v_1-|E_{0s}|^2\Bigg(v_1
2u_0v_1+2v_0u_1+\frac{2}{\sqrt{3}}u_1v_1+\frac{1}{\sqrt{3}}v_0u_1^2
\nonumber\\
&+&2u_0v_0u_1+u_0^2v_1+\frac{5}{4}u_1^2v_1
+\frac{2}{\sqrt{3}}u_0u_1v_1
+3v_0^2v_1+\sqrt{3}v_0v_1^2
+\frac{5}{4}v_1^3\Bigg)\,,
\\
0&=&-v_1-2u_1+|E_{0s}|^2\Bigg(
3u_1+6u_0u_1+\sqrt{3}u_1^2+2v_0v_1+\frac{1}{\sqrt{3}}v_1^2
\nonumber\\
&+&3u_0^2u_1+\sqrt{3}u_0u_1^2+\frac{5}{4}u_1^3+2u_0v_0v_1
+\frac{1}{\sqrt{3}}u_0v_1^2
+v_0^2u_1+\frac{2}{\sqrt{3}}v_0u_1v_1
+\frac{5}{4}u_1v_1^2\Bigg)\,.
\end{eqnarray}
The above equations are solved numerically with the help 
of Mathematica,
 by using, for simplicity, the 
condition $\Delta=|E_{0s}|^2$, which assures that $E_{in} =E_{0s}$ 
(see Eq. (\ref{E0s}). 
The solutions are compared to the steady-state 
mode amplitudes obtained by numerically
integrating the dynamical equations (\ref{cdyn}); 
a perfect agreement between 
the two approaches has been found.

\section{Figures}

\end{document}